# Stability of Bloch Oscillations in two coupled Bose-Einstein condensates

Vikash Ranjan[a], Aranya B Bhattacherjee[b] and ManMohan[a*]

[a]Department of Physics and Astrophysics, University of Delhi, Delhi-110 007, India

[b]Department of Physics, A.R.S.D College, University of Delhi (South Campus), Dhaula Kuan, New Delh-110 021, India

**Abstract**: We investigate analytically, the stability of Bloch waves at the boundary of the first Brillouin zone in two coupled Bose-Einstein condensates confined in an optical lattice. Contrary to the single component case, we find here two critical density regimes which determine the stability of the Bloch waves. Breakdown of Bloch oscillations appear when $n_1/n_2 < N_{C1}$ or when $n_1/n_2 > N_{C2}$, here $N_{C1}$ and $N_{C2}$ are some critical values of $n_1/n_2$. There is an intermediate regime between $N_{C1}$ and $N_{C2}$ where the Bloch oscillations are stable and the condensates behave like single particles.

* To whom correspondence should be addressed (sneh@del2.vsnl.net.in)

The experimental realization of confining Bose-Einstein condensates (BEC) in optical lattices has opened new opportunities to study and probe concepts developed for condensed matter at totally different scales. These include Josephson effect [1], Bloch oscillations [2] superfluid to Mott insulator transition [3] and most recently Tonks-Girardeau gas[4]. In the context of "Stability of Bloch waves in optical lattices", it has been predicted that the chemical potential of the condensate becomes multi-valued near the zone boundary [5-7] and hence indicating the onset of instability. In particular Diakonov et al.[7] showed that the instability of the Bloch waves occurs at densities greater than a certain critical density. Motivated by this interesting and peculiar behavior of a BEC near the zone boundary and the outburst of some interesting physics associated with two condensates [8], we naturally asked the question "How would the dynamics of Bloch oscillations of a condensate change in the presence of a second one?" .To this end, in this article, we look into the stability of Bloch waves at the boundary of the first Brillouin zone in two coupled Bose-Einstein condensates confined in an optical lattice. We will closely follow the formalism developed in reference [7].

We consider two BEC's confined by one dimensional periodic potentials of the form

$$V_1 = V_{01} Sin^2 \left( \frac{\pi x}{d} \right) \tag{1}$$

$$V_2 = V_{02} Cos^2 \left( \frac{\pi x}{d} \right) \tag{2}$$

Where $V_{01}$ and $V_{02}$ are the heights of the potentials. $d$ is the period of the lattices. These potentials are created by the superposition of two counter-propagating orthogonal plane polarized waves giving rise to a standing wave for which the polarization is space dependent. The polarization is alternatively circular $\sigma^-$ in $x = 0$, linear in $x = \lambda/8$, circular $\sigma^+$ in $x = \lambda/4$ etc. with a spatial periodicity $\lambda/2$. This polarization gradient leads to a spatial modulation of the degenerate ground state Zeeman sublevels, which act as a potential for the motion of the atomic centre of mass. The light shifts of the Zeeman sub-levels are position dependent, since the Clebsch-Gordan couplings vary with polarization. The electric field propagating along the $x$-axis is

$$\vec{E}(x,t) = \frac{E_o}{\sqrt{2}} \left\{ \vec{\varepsilon}_- e^{i\pi/4} \cos(k_L x) + \vec{\varepsilon}_+ e^{-i\pi/4} \sin(k_L x) \right\}, \text{ where } \vec{\varepsilon}_\pm = m(\vec{e}_x \pm i\vec{e}_y)/\sqrt{2}$$

are the polarization unit vectors corresponding to $\sigma^\pm$ polarizations, $E_o$ is the amplitude and $k_L$ is the wave vector of the lasers. The detuning of both the waves must be large so that Raman terms can be neglected. For actual experiments, we can use either two isotopes of the same atom (e.g. $^{87}$Rb atoms in hyperfine states F=2, m$_F$=2 and F=1, m$_F$=-1) or two different atoms.

The two BEC's $\psi_1$ and $\psi_2$ obey the coupled Gross-Pitaevskii equations

$$\frac{-\hbar^2}{2m_1}\nabla^2\psi_1 + V_1\psi_1 + U_{11}|\psi_1|^1\psi_1 + U_{12}|\psi_2|^2\psi_1 = \mu_1\psi_1 \tag{3}$$

$$\frac{-\hbar^2}{2m_2}\nabla^2\psi_2 + V_2\psi_2 + U_{22}|\psi_2|^1\psi_2 + U_{12}|\psi_1|^2\psi_2 = \mu_2\psi_2 \tag{4}$$

$\mu_1$ and $\mu_2$ are the chemical potentials of the two BEC's and $U_{ij}$ is the two body interaction and is related to the s-wave scattering length $a_{ij}$ by $U_{ij} = 2\pi\hbar^2 a_{ij}/m_{ij}$, $m_{ij} = m_{ij}/(m_i + m_j)$. $m_i$ is the particle mass.

The average total energy of the two condensates over a lattice period corresponding to equations (3) and (4) is

$$E = \frac{1}{d}\int_{-d/2}^{d/2} dx \left[ \frac{\hbar^2}{2m_1}\left|\frac{d\psi_1}{dx}\right|^2 + V_1(x)|\psi_1|^2 + \frac{\hbar^2}{2m_2}\left|\frac{d\psi_2}{dx}\right|^2 + V_2(x)|\psi_2|^2 + \frac{1}{2}U_{11}|\psi_1|^4 + \frac{1}{2}U_{22}|\psi_2|^4 + U_{12}|\psi_1|^2|\psi_2|^2 \right] \tag{5}$$

The mean number densities $n_1$ and $n_2$ are given by

$$n_i = \frac{1}{d}\int_{-d/2}^{d/2} dx |\psi_i|^2, \quad i=1,2 \tag{6}$$

We shall look for solutions where the mean density is independent of position and therefore these must satisfy the condition for quasi-periodicity for all $x$.

$$\psi(x+d) = \exp(ikd)\psi(x) \tag{7}$$

The quantity $\hbar k$ is the quasi-momentum of the condensate. The structure of the energy at the zone boundary can be found by making a variational calculation of the energy. We use trial functions of the form

$$\psi_1(x) = \sqrt{n_1}\left\{\cos(\alpha)\exp(ikx) + \sin(\alpha)\exp i(k - 2\pi/d)x\right\} \tag{8a}$$

$$\psi_2(x) = \sqrt{n_2}\left\{\sin(\alpha)\exp(ikx) - \cos(\alpha)\exp i(k - 2\pi/d)x\right\} \tag{8b}$$

Where $\alpha$ is a variational parameter. The energy functional (5) evaluated for the trial wave-functions (8) is given by

$$E = n_1\left\{\frac{(k-2\pi/d)^2}{4m_1} + \frac{k^2\hbar^2}{4m_1} + 2\left(k-\frac{\pi}{d}\right)\frac{\pi\hbar^2}{2m_1 d^2}\cos(\alpha)\right\} + \frac{n_1 V_{01}}{2}\left(1 - \frac{\sin(2\alpha)}{2}\right)$$
$$+ \frac{n_1^2 U_{11}}{2}\left(1 + \frac{\sin^2(2\alpha)}{2}\right) + n_2\left\{\frac{(k-2\pi/d)^2}{4m_2} + \frac{k^2\hbar^2}{4m_2} + 2\left(k-\frac{\pi}{d}\right)\frac{\pi\hbar^2}{2m_2 d^2}\cos(\alpha)\right\} \quad (9)$$
$$\frac{n_2 V_{02}}{2}\left(1 - \frac{\sin(2\alpha)}{2}\right) + \frac{n_2^2 U_{22}}{2}\left(1 + \frac{\sin^2(2\alpha)}{2}\right) + n_1 n_2 U_{12}\left(1 - \frac{\sin^2(2\alpha)}{2}\right)$$

The optimal value of $\alpha$ is obtained by requiring eqn.(9) to be stationary with respect to variations in $\alpha$, and the chemical potential $\mu_i$ is then found by differentiating the resulting energy with respect to the density, $\mu_i = \partial E / \partial n_i$, $i = 1,2$. At the zone boundary $k = \pm \pi / d$ the stationarity of eqn.(9) yields two solutions

$$\cos(2\alpha) = 0 \tag{10}$$

And

$$\sin(2\alpha) = \frac{(n_1 V_{01} + n_2 V_{02})}{2(n_1^2 U_{11} + n_2^2 U_{22} - 2n_1 n_2 U_{12})} \tag{11}$$

The above two stationary solutions yield two values of the energies. The first solution yields

$$E^+ = n_1 E_{r1} + n_2 E_{r2} + \frac{1}{4}(n_1 V_{01} + n_2 V_{02}) + \frac{3}{4}(n_1^2 U_{11} + n_2^2 U_{22}) + \frac{1}{2} n_1 n_2 U_{12} \tag{12}$$

While the second solution yields

$$E^- = \frac{\pi \hbar^2}{2d^2}\left[\frac{n_1}{m_1} + \frac{n_2}{m_2}\right] + \frac{1}{2}[n_1 V_{01} + n_2 V_{02}] + \frac{1}{2}[n_1^2 U_{11} + n_2^2 U_{22} + 2n_1 n_2 U_{12}]$$
$$- \frac{(n_1 V_{01} + n_2 V_{02})^2}{16(n_1^2 U_{11} + n_2^2 U_{22} - 2n_1 n_2 U_{12})} \tag{13}$$

$E_{ri} = \pi^2 \hbar^2 / 2 m_i d^2$ are the recoil energies of the two components.

Eqns. (12) and (13) yields four values of the chemical potential for the two components

$$\mu_1^+ = \frac{\partial E^+}{\partial n_1} = E_{r1} + \frac{V_{01}}{4} + \frac{3}{2}n_1 U_{11} + \frac{1}{2}n_2 U_{12} \qquad (14)$$

$$\mu_2^+ = \frac{\partial E^+}{\partial n_2} = E_{r2} + \frac{V_{02}}{4} + \frac{3}{2}n_2 U_{22} + \frac{1}{2}n_1 U_{12} \qquad (15)$$

$$\mu_1^- = \frac{\partial E^-}{\partial n_1} = E_{r1} + \frac{V_{01}}{2} + n_1 U_{11} + n_2 U_{12}$$
$$- n_2(n_1 V_{01} + n_2 V_{02}) \left[ \frac{\{V_{01}(n_2 U_{22} - n_1 U_{12}) - V_{02}(n_1 U_{11} - n_2 U_{12})\}}{8(n_1^2 U_{11} + n_2^2 U_{22} - 2n_1 n_2 U_{12})^2} \right] \qquad (16)$$

$$\mu_2^- = \frac{\partial E^-}{\partial n_2} = E_{r2} + \frac{V_{02}}{2} + n_2 U_{22} + n_1 U_{12}$$
$$- n_1(n_1 V_{01} + n_2 V_{02}) \left[ \frac{\{V_{02}(n_1 U_{11} - n_2 U_{12}) - V_{01}(n_2 U_{22} - n_1 U_{12})\}}{8(n_1^2 U_{11} + n_2^2 U_{22} - 2n_1 n_2 U_{12})^2} \right] \qquad (17)$$

Bloch oscillations of a BEC confined in a periodic potential is a manifestation of the quantum dynamics of the condensate. On applying a force, the condensate moves through the Brillouin zone until they reach the band edge, where they encounter a gap between the first and the second band. In the absence of sufficient acceleration, the BEC is unable to make a Landau-Zener (LZ) transition to the upper band and are instead Bragg reflected back to the opposite end of the Brillouin zone [2]. These correspond to periodic oscillations of the condensate velocity and are termed as Bloch oscillations. In the laboratory frame these Bloch oscillations manifest themselves as an undulating increase in velocity rather than the linear increase expected in the classical picture.

Diakonov et al. had shown that at low densities of the condensate, the dependence of the energy on the quasi-momentum is similar to that for a single particle, but at densities greater than a critical one the chemical potential of the condensate becomes multi-valued near the boundary of the first Brillouin zone and develops a characteristic loop structure. A direct consequence of this loop structure is the breakdown of the Bloch oscillations due to the non-zero adiabatic tunneling into the upper band. This breakdown of Bloch oscillations in the present system of two condensates is studied using eqns (10), (11) and (14-17). In particular we will concentrate only on the dynamics of the system at the boundary of the first Brillouin where the interesting physics occurs.

Figure 1 show a plot of $\mu_1^{+,-}/E_{r1}$ (at the zone boundary) as a function of $n_1/n_2$ ($n_2$ is held fixed while $n_1$ is varied continuously). The dual values of the chemical potential indicates the presence of the loop structure and hence breakdown of the Bloch oscillations. Contrary to the single component case, we find here two critical density regimes which determine the stability of the Bloch waves. Breakdown of Bloch oscillations appear when $n_1/n_2 < N_{C1}$ or when $n_1/n_2 > N_{C2}$, where $N_{C1}$ and $N_{C2}$ are some critical values of $n_1/n_2$. There is an intermediate regime between $N_{C1}$ and $N_{C2}$ where the Bloch oscillations are stable and the condensate behaves like a single particle. This happens because in the intermediate regime, the condensate does not have sufficient energy to make a LZ transition to the upper band and is Bragg reflected instead.

Fig.2 shows a plot of $\mu_2^{+,-}/\mu_1^{+,-}$ as a function of $n_2/n_1$ for two different values of $n_1$ while $n_2$ is varied continuously. The stable region appears to increase with decreasing value of $n_1$. In the absence of the first component (i.e. $n_1=0$), $N_{C1}$ vanishes (not shown in this figure) and the loop structure appears only for $n_2 > n_{cr}$ ($n_{cr}$ is the critical density defined in ref. [7]). For $n_1/n_2 < N_{C1}$ or $n_1/n_2 > N_{C2}$, $\sin(2\alpha) < 1$ and the solutions 8(a) and 8(b) are complex. At the Brillouin zone ($k = \pm \pi/d$) there is a second solution equal to the complex conjugate of the first one and in addition, a third real solution corresponding to $\cos(2\alpha) = 0$. In the intermediate regime, $\sin(2\alpha) > 1$ and hence there is only one stable solution. A plot of $\sin(2\alpha)$ as a function of $n_1/n_2$ for three different values of $n_{cr}/n_2$ is shown in fig.3. For simplicity we have taken $V_{01} = V_{02} = V_0$ and $U_{11} = U_{22} \approx U_{12} = U_0$, corresponding to two different internal states of the same atom. $n_{cr} = V_0/2U_0$. Clearly the stability region ($\sin(2\alpha) > 1$) decreases with decreasing $n_{cr}/n_2$.

In conclusion, we find that stable Bloch oscillations of a system of two condensates are restricted to a narrow density region between a lower and a higher limit. This is in contrast to the case of a single BEC, where the lower limit does not exist. A possible explanation for such a behavior could be that, the presence of the second condensate lessens the single particle behavior (hence the stable Bloch oscillations) of the first condensate due to the two body interaction.

**Figure 1**

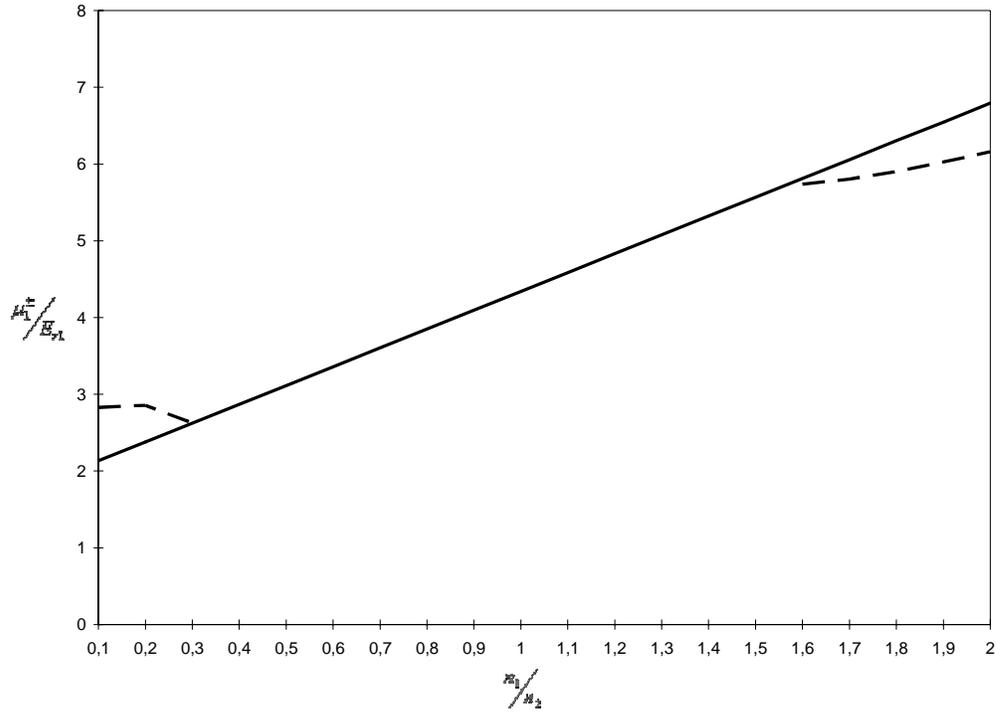

Figure 1: A plot of $\mu_1^{+,-}/E_{r1}$ (at the zone boundary) as a function of $n_1/n_2$ ($n_2$ is held fixed while $n_1$ is varied continuously). The first component is $^{23}$Na and the second component is $^{87}$Rb. $U_{11}/U_{22} = 1.82$, $U_{12}/U_{22} = 1.42$, $V_{01} = E_{r1}$ and $V_{02} = E_{r2}$.

**Figure 2**

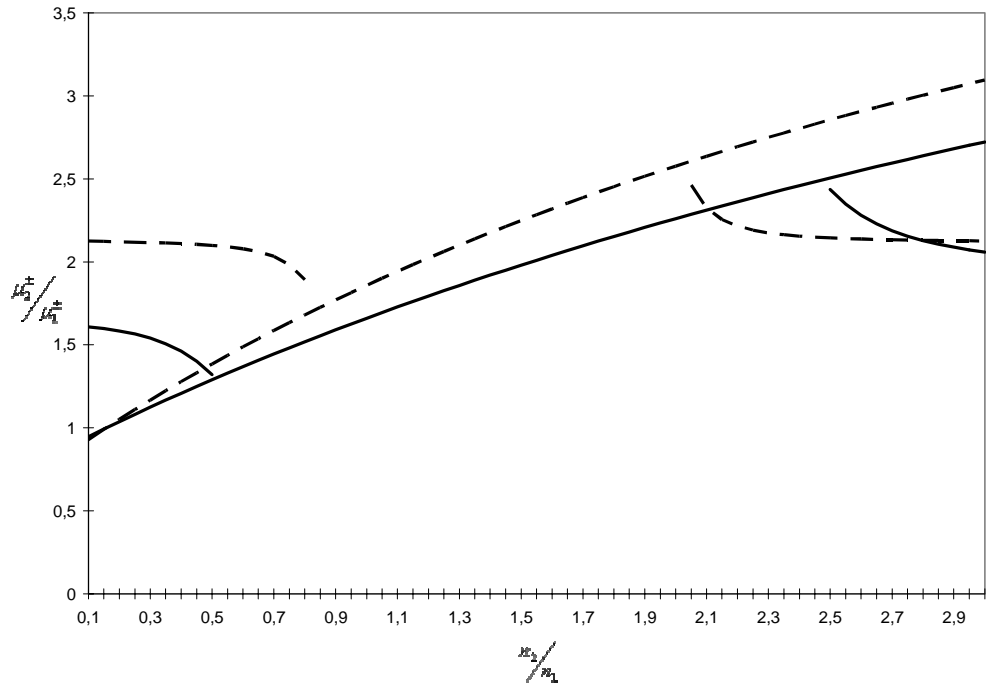

Figure 2: A plot of $\mu_2^{+,-}/\mu_1^{+,-}$ as a function of $n_2/n_1$ for two different values of $n_1$ while $n_2$ is varied continuously. Other data are same as in figure 1. The dashed line is for $n_1 = 5.0 \times 10^{21}$ and the continuous dark line is for $n_1 = 1.0 \times 10^{21}$.

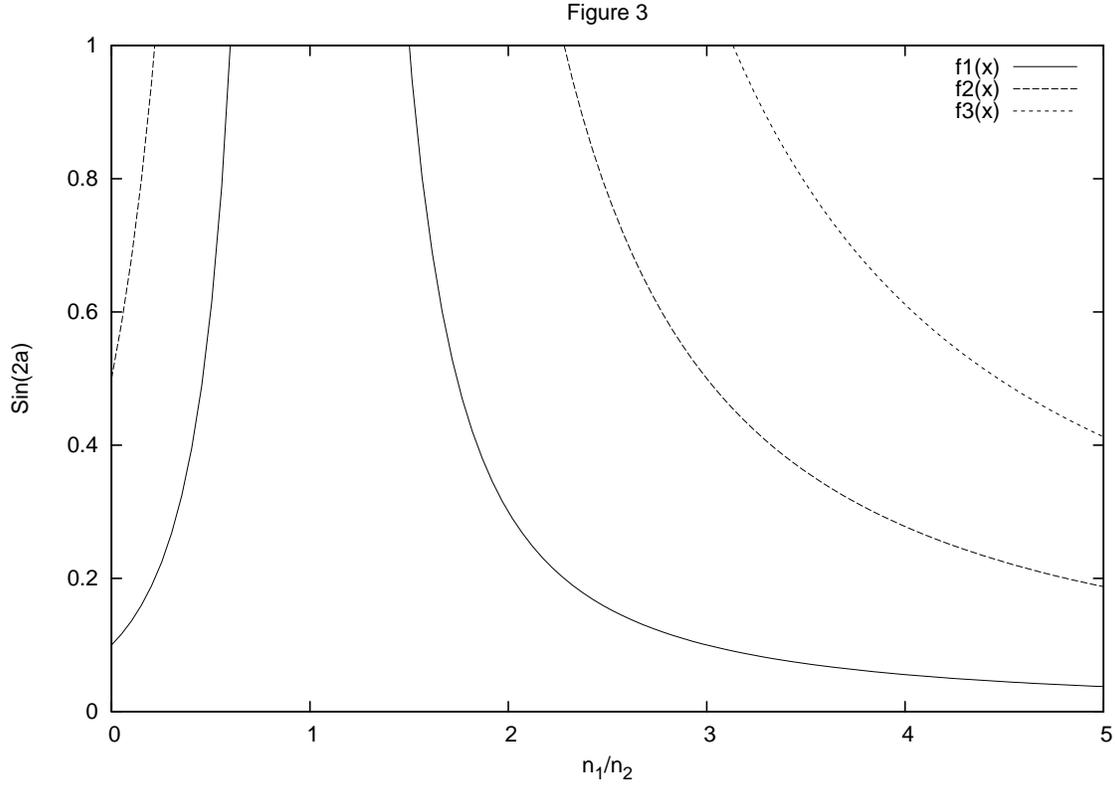

Figure 3: A plot of $\sin(2\alpha)$ as a function of $n_1/n_2$ for three different values of $n_{cr}/n_2$. for simplicity we have taken $V_{01} = V_{02} = V_0$ and $U_{11} = U_{22} \approx U_{12} = U_0$, corresponding to two different internal states of the same atom and $n_{cr} = V_0/2U_0$. $n_{cr}/n_2 = 1.1$ (short dashed line), $n_{cr}/n_2 = 0.5$ (dashed line) and $n_{cr}/n_2 = 0.1$ (continuous dark line).